\documentclass[aps,prl,twocolumn,preprintnumbers,10pt,showpacs,nohyper]{revtex4}
\usepackage{amsmath,bm}

\usepackage{graphicx}
\usepackage{epsfig}

\def\dsigma{{\rm d} \hat\sigma}

\allowdisplaybreaks

\begin{document}

\preprint{IPPP/13/3, ZU-TH 03/13}

\title{Second order QCD corrections to jet production at 
hadron colliders:\\ the all-gluon contribution}

\author{A.\ Gehrmann-De Ridder$^a$, T.\ Gehrmann$^b$, 
E.W.N.\ Glover$^c$, J.\ Pires$^a$}
\affiliation{$^a$ Institute for Theoretical Physics, ETH, 
CH-8093 Z\"urich, Switzerland\\
$^b$ Institute for Theoretical Physics, University of
  Z\"urich, CH-8057 Z\"urich, Switzerland\\
$^c$ Institute for Particle Physics 
Phenomenology, University of Durham, Durham DH1 3LE, England}


\begin{abstract}
We report the calculation of next-to-next-to-leading order (NNLO) 
QCD corrections in the  purely gluonic channel
to dijet production and related observables at hadron 
colliders. Our result represents the first  NNLO calculation of 
a massless jet observable at hadron colliders, and opens the path towards
precision QCD phenomenology with the LHC. 
\end{abstract}

\maketitle

Single inclusive jet and dijet observables are the most 
fundamental QCD 
processes measured at hadron colliders. They probe the basic 
parton-parton scattering in $2\to 2$ kinematics, and thus allow for 
a determination of the parton distribution functions in the proton 
and for a direct probe of the strong coupling constant $\alpha_s$ up 
to the highest energy scales that can be attained in collider 
experiments. 

Precision measurements of single jet and dijet cross sections
have been performed by CDF~\cite{cdfjet} and D0~\cite{d0jet} 
at the Tevatron and
by ATLAS~\cite{atlasjet} and CMS~\cite{cmsjet} 
at the LHC. The Tevatron data are included in nearly all 
global fits of parton distributions, where they provide crucial 
information on the gluon content of the proton, and have been  
used to determine the strong coupling constant~\cite{asjet}.

Theoretical predictions for  these observables are accurate to 
next-to-leading order (NLO) in 
QCD~\cite{eks,jetrad,nlojet1,powheg2j,meks} and the 
electroweak theory~\cite{Dittmaier:2012kx}.
The estimated uncertainty from missing higher order corrections 
on the NLO QCD predictions is substantially larger than the 
experimental errors on single jet and dijet data, and is thus the dominant 
source of error in the determination of $\alpha_s$. A consistent inclusion of 
jet data in global fits of parton distributions is also  feasible only to 
NLO.  These theoretical limitations to precision phenomenology provide a
very strong motivation for computing next-to-next-to-leading order (NNLO)
corrections to jet production at hadron colliders. 

At this perturbative order, three types of parton-level 
processes contribute to jet production: the two-loop 
virtual corrections to 
the basic $2\to 2$ process~\cite{twol}, the one-loop 
virtual corrections 
to the single real radiation $2\to 3$ process~\cite{onelv} and the 
double real radiation $2\to 4$ process at tree-level~\cite{real}. 
Each contribution 
is infrared divergent, and only their sum yields a finite and  
meaningful result.   After 
ultraviolet renormalization, both virtual contributions 
contain explicit infrared singularities, which are compensated 
by infrared singularities from single or double real radiation. 
These become explicit only after integrating out the real 
radiation contributions over the phase space relevant to 
single jet or dijet production. This interplay with the jet definition
complicates the extraction of infrared singularities 
from the real radiation process. It is typically done by subtracting 
an infrared approximation from the corresponding matrix elements. These 
infrared subtraction terms are sufficiently simple to be integrated 
analytically, such that they can be combined with the virtual contributions 
to obtain the cancellation of all infrared singularities. 
Several generic methods
for the construction of subtraction terms are available at NLO~\cite{fks,cs,ant}.

The development of subtraction methods for NNLO calculations is 
a very active field of research. Up to now, various methods were constructed 
and applied to specific NNLO calculations of exclusive observables:
sector decomposition~\cite{secdec} applied to Higgs
production~\cite{babishiggs} and vector boson
production~\cite{kirilldy};  $q_T$-subtraction~\cite{qtsub} 
to Higgs production~\cite{grazzinihiggs}, vector boson
production~\cite{grazzinidy}, associated
$VH$-production~\cite{grazziniwh}, photon pair production~\cite{grazzinigg}
and  top quark decay~\cite{scettop}; antenna subtraction~\cite{ourant}
to three-jet production~\cite{our3j,weinzierl3j} and related event 
shapes~\cite{ourev,weinzierlev} in 
 $e^+e^-$ annihilation; and sector-improved residue 
subtraction~\cite{stripper} to top quark pair production~\cite{czakontop}. 

The antenna subtraction method~\cite{ourant,hadant,currie1} 
constructs subtraction terms 
from antenna functions which encapsulate all unresolved radiation in between 
a pair of hard radiator partons. At NNLO,  antenna functions with up 
to two unresolved partons at tree level and one unresolved parton 
at one loop are required. For hadron collider observables, 
one~\cite{gionata} or both~\cite{ritzmann,monni} radiator partons can be in the initial state.  

For  the NNLO all-gluon contribution to jet production at hadron colliders, 
the antenna subtraction terms were constructed for the 
tree-level double real 
radiation process in~\cite{joao1} and for the one-loop single real radiation 
process in~\cite{joao2}. These subtraction terms were integrated and 
combined~\cite{joao3} 
with the relevant two-loop matrix elements and parton 
distributions, resulting 
in a full cancellation of infrared poles. We have now implemented 
these terms into a parton-level event generator, which can compute 
the all-gluon contribution to any
infrared-safe  observable related to dijet final states at hadron 
colliders to NNLO accuracy. The program consists of three integration
channels:
\begin{eqnarray}
\dsigma_{gg,NNLO}&=&\int_{{\rm{d}}\Phi_{4}}\left[\dsigma_{gg,NNLO}^{RR}-\dsigma_{gg,NNLO}^S\right]
\nonumber \\
&+& \int_{{\rm{d}}\Phi_{3}}
\left[
\dsigma_{gg,NNLO}^{RV}-\dsigma_{gg,NNLO}^{T}
\right] \nonumber \\
&+&\int_{{\rm{d}}\Phi_{2}}\left[
\dsigma_{gg,NNLO}^{VV}-\dsigma_{gg,NNLO}^{U}\right],
\end{eqnarray}
where each of the square brackets is finite and well 
behaved in the infrared singular regions. For the all-gluons 
channel, the construction of 
the three subtraction terms $\dsigma_{ij,NNLO}^{S,T,U}$ was described in 
Refs.~\cite{joao1,joao2,joao3}. 

In the three-parton and four-parton channel, the phase space has  been 
decomposed into multiple wedges (6 three-parton wedges and 
30 four-parton wedges), each containing only a subset of 
possible infrared singular contributions. Inside each wedge, the generation 
of multiple phase space configurations related by angular rotation 
of unresolved pairs of particles around their common momentum axis 
ensures a local convergence of the antenna subtraction term to the 
relevant matrix element. Owing to the symmetry properties of the all-gluon 
final state, many wedges yield identical contributions, thereby allowing 
a substantial speed-up of their evaluation. 

Jets in hadronic collisions can be produced through a variety of different 
partonic subprocesses, and the all-gluon process is only one of them.  Our
results on this process can therefore not be directly compared  with
experimental data. The all-gluon process does however allow to establish the
calculational method, and to qualify the  potential impact of NNLO corrections
on jet observables. It should be  noted that the NLO corrections to hadronic
two- and three-jet production were  also first derived in the all-gluon
channel~\cite{dijetgonly,trocsanyi,giele}, well  before full results could be
completed~\cite{eks,jetrad,nagy}. In both cases,  the all-gluon results were
extremely vital both for establishing the methodology and for assessing the
infrared sensitivity of different jet algorithms~\cite{giele}.

Our numerical studies for proton-proton collisions at 
centre-of-mass energy $\sqrt{s}=8$~TeV
concern the single jet inclusive cross section 
(where every identified jet in an event that passes the selection cuts 
contributes, such that 
a single event potentially enters the distributions multiple times) 
and the two-jet exclusive cross section (where events
with exactly two identified jets contribute). 

Jets are identified using the anti-$k_T$ algorithm with resolution 
parameter $R=0.7$. Jets are accepted at central rapidity $|y|<4.4$, and 
ordered in transverse momentum. An event is retained if the leading 
jet has $p_{T1}>80$~GeV.  For the dijet invariant mass distribution, a second jet must be observed with $p_{T2}>60$~GeV.

\begin{figure}[th]
  \centering
    \includegraphics[width=0.5\textwidth]{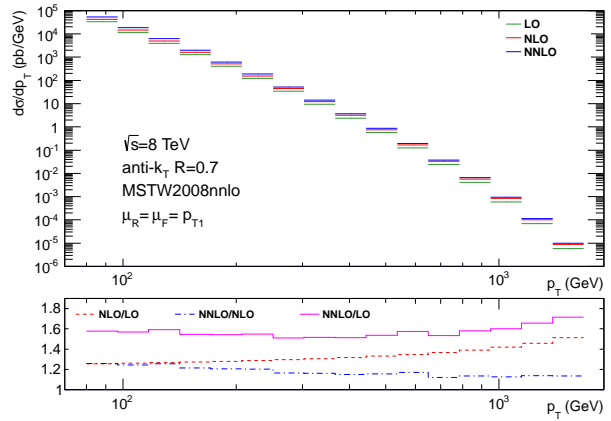}
  \caption{Inclusive jet transverse energy distribution, $d\sigma/dp_T$, for jets constructed with the anti-$k_T$ algorithm with $R=0.7$ and with $p_T > 80$~GeV, $|y| < 4.4$ and $\sqrt{s} = 8$~TeV at NNLO (blue), NLO (red) and LO (dark-green). The lower panel shows the ratios of NNLO, NLO and LO cross sections.}
  \label{fig:dsdet}
\end{figure}

All calculations are carried out with the MSTW08NNLO gluon 
distribution function~\cite{mstw}, including the evaluation of the 
LO and NLO contributions \footnote{Note that the evolution of the gluon distribution within the PDF set together with the value of $\alpha_s$ intrinsically includes contributions from the light quarks.  The NNLO calculation presented here is ``gluons-only" in the sense that only gluonic matrix elements are involved.}.
This choice of parameters
allows us to quantify the size of the genuine NNLO contributions to the  
parton-level subprocess. Factorization and renormalization scales
($\mu_F$ and $\mu_R$)   
are chosen dynamically on an event-by-event basis. As default value, we
set $\mu_F = \mu_R \equiv \mu $ and set $\mu$ equal to the transverse momentum of the leading jet so that $\mu = p_{T1}$. 

\begin{figure}[th]
  \centering
    \includegraphics[width=0.5\textwidth]{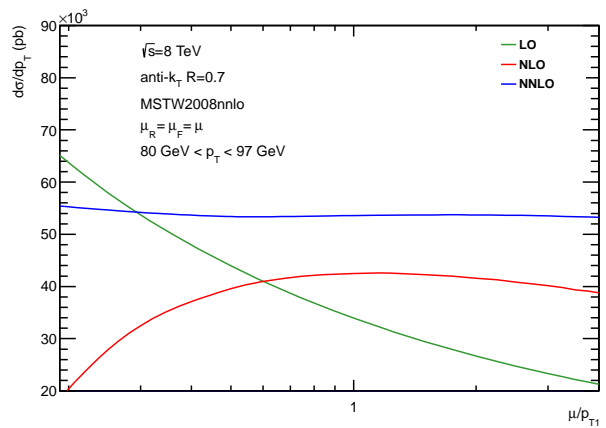}
  \caption{Scale dependence of the inclusive jet cross section for $pp$ collisions at $\sqrt{s}=8~$TeV for the anti-$k_T$ algorithm with $R=0.7$ and with $|y| < 4.4$ and $80$~GeV $< p_T < 97$~GeV at NNLO (blue), NLO (red) and LO (green).}
  \label{fig:dsdmu}
\end{figure}

In Fig.~\ref{fig:dsdet} we present the inclusive jet cross section for the
anti-$k_T$ algorithm with $R=0.7$ and with $p_T > 80$~GeV, $|y| < 4.4$ as a
function of the jet $p_{T}$ at LO, NLO and NNLO, for the central scale choice
$\mu = p_{T1}$.  The NNLO/NLO $k$-factor shows the size of the higher order NNLO
effect to  the cross section in each  bin with respect to the NLO calculation.
For this scale choice we see that the NNLO/NLO $k$-factor is approximately flat
across the $p_{T}$ range corresponding to a 15-25\% increase compared to the NLO
cross section.

One of the main motivations for computing the NNLO QCD corrections is to reduce the scale uncertainty in the theoretical prediction.  This is illustrated in Fig.~\ref{fig:dsdmu} for the single jet inclusive cross section for
jets with $|y| < 4.4$ and $80$~GeV $< p_T < 97$~GeV.  We see that 
the scale dependence of the cross section at NNLO is vastly reduced.   The scale dependence of other $p_T$ and $y$ slices is also reduced.

\begin{figure}[th]
  \centering
    \includegraphics[width=0.51\textwidth]{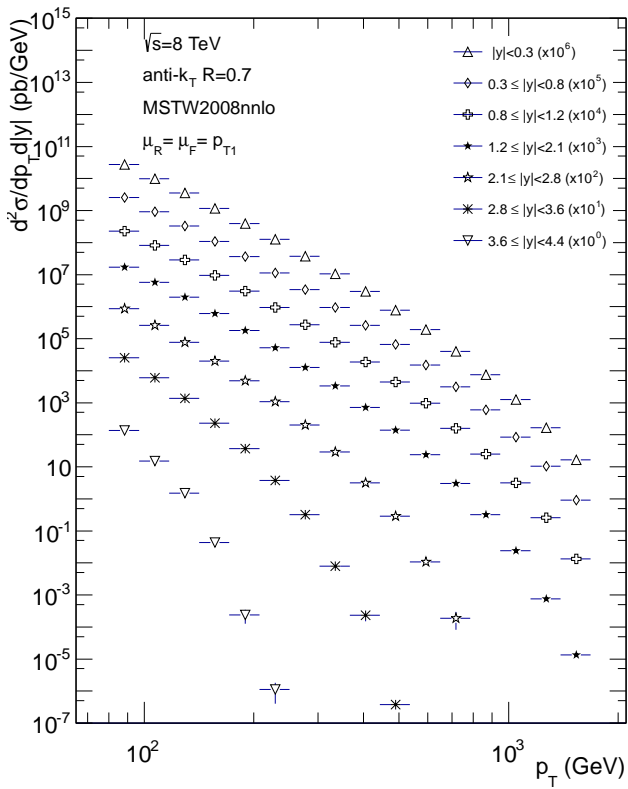}
  \caption{The doubly differential inclusive jet transverse energy distribution, $d^2\sigma/dp_T d|y|$, at $\sqrt{s} = 8$~TeV for the anti-$k_T$ algorithm with $R=0.7$ and for $E_T > 80$~GeV and various $|y|$ slices.}
  \label{fig:d2sdetslice}
\end{figure}

To illustrate the range of observables that can be studied with our computation we show in Fig.~\ref{fig:d2sdetslice} the inclusive jet cross section in double-differential form in
jet $p_{T}$ and rapidity bins at NNLO. 
The $p_{T}$ range is divided into 16 jet-$p_{T}$ bins and seven rapidity intervals over the range 0.0-4.4 covering
central and forward jets. 

\begin{figure}[th]
  \centering
    \includegraphics[width=0.5\textwidth]{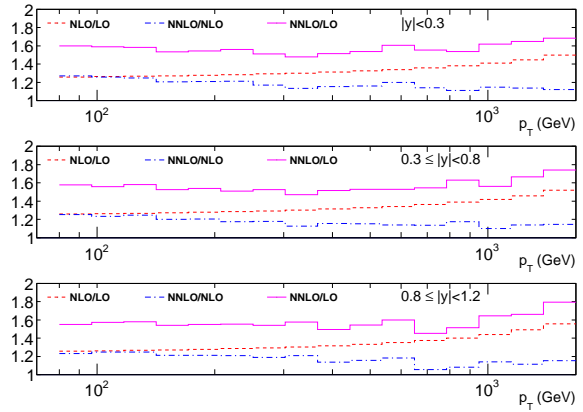}
  \caption{Double differential $k$-factors for $p_T > 80$~GeV and three $|y|$ slices: $|y | < 0.3$, $0.3 < |y| < 0.8$ and $0.8 < |y| < 1.2$.}
  \label{fig:d2kdetslice}
\end{figure}

Fig.~\ref{fig:d2kdetslice} shows the double-differential $k$-factors for the distribution in Fig.~\ref{fig:d2sdetslice} for three rapidity slices: $|y | < 0.3$, $0.3 < |y| < 0.8$ and $0.8 < |y| < 1.2$.
We observe that the NNLO correction increases the cross section between 25\% at low $p_{T}$ to 12\% at high $p_{T}$ with respect to the NLO calculation
and this behaviour is similar for all three rapidity slices.

\begin{figure}[th]
  \centering
    \includegraphics[width=0.5\textwidth]{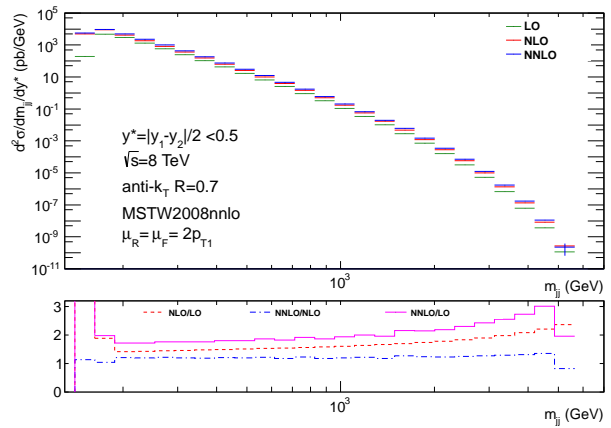}
  \caption{Exclusive dijet invariant mass distribution, $d\sigma/dm_{jj}dy^*$, at $\sqrt{s} = 8$~TeV for $y^* < 0.5$ with $p_{T1} > 80$~GeV, $p_{T2} > 60$~GeV and $|y_1|,~|y_2| < 4.4$ at NNLO (blue), NLO (red) and LO (dark-green). The lower panel shows the ratios of NNLO, NLO and LO cross sections.}
  \label{fig:dsdmjj}
\end{figure}

As a final observable, we computed the dijet cross section as a function of the
dijet mass at NNLO. This is shown in Fig.~\ref{fig:dsdmjj} for the scale choice
$\mu = 2p_{T1}$ together with the LO and NLO results. The dijet mass is computed
from the two jets with the highest $p_{T}$ and $|y_1|,~|y_2| < 4.4$ with $y^{*}$, defined as half the
rapidity difference of the two leading jets $y^{*}=|y_{1}-y_{2}|/2<0.5$.  We see
that the NNLO/NLO $k$-factor is approximately flat across the $m_{jj}$ range
corresponding to a 15-20\% increase compared to the NLO cross section.

In conclusion, we have described the first calculation of the fully differential
inclusive jet and dijet cross sections  at hadron colliders at NNLO in the
strong coupling constant using the new parton-level generator NNLOJET. We have
considered the NNLO QCD corrections from the purely gluonic channel at leading
colour. As demonstrated in~\cite{joao2,joao3}, using the antenna subtraction
scheme the explicit $\epsilon$-poles in the dimension regularization parameter
of one- and two-loop matrix elements entering this calculation are cancelled in
analytic and local form against the $\epsilon$-poles of the integrated antenna
subtraction terms thereby enabling the computation of jet cross sections at
hadron colliders at NNLO accuracy.  All of these techniques can be readily
applied to the quark contributions.

For all of the observables considered here, we observed a dramatic reduction of the respective
uncertainties in the theory prediction due to variations of the factorization
and renormalization scales. We expect similar conclusions when including the
processes involving quarks.

This research was supported in part by the UK Science and Technology Facilities
Council,  in part by the Swiss National Science Foundation (SNF) under contracts
PP00P2-139192 and 200020-138206 and in part by the European Commission through the ``LHCPhenoNet"
Initial Training Network PITN-GA-2010-264564. EWNG gratefully acknowledges the
support of the Wolfson Foundation, the Royal Society and the Pauli Center for Theoretical Studies.

\end{document}